\documentclass[pra, a4paper, preprint, showpacs,superscriptaddress]{revtex4-2}
\usepackage{amsmath,amscd,amsfonts,amssymb, amsbsy, color,bbm, bm,amsthm}
\usepackage{graphicx}
\usepackage[T1]{fontenc}
\usepackage{enumitem}
\newcommand{\be}{\begin{equation}}
\newcommand{\ee}{\end{equation}}

\newcommand{\ba}[1]{\left(\begin{array}{#1}}
\newcommand{\ea}{\end{array}\right)}

\theoremstyle{plain}  
\newtheorem{thm}{Theorem}

\newtheorem*{pf}{Proof}

\begin{document}
\title{Lorentz invariants of pure three-qubit states}  

\author{A. R. Usha Devi} 
\affiliation{Department of Physics, Bangalore University, 
	Bangalore-560 056, India}
\affiliation{Inspire Institute Inc., Alexandria, Virginia, 22303, USA.}
\email{ushadevi@bub.ernet.in}

\author{Sudha} 
\affiliation{Department of Physics, Kuvempu University, 
	Shankaraghatta-577 451, Karnataka, India}
\email{tthdrs@gmail.com}
\affiliation{Inspire Institute Inc., Alexandria, Virginia, 22303, USA.}

\author{H. Akshata Shenoy} 
\affiliation{International Centre for Theory of Quantum Technologies, University of Gd{\'a}nsk, Gd{\'a}nsk, Poland}
\email{akshata.shenoy@ug.edu.pl}

\author{H. S. Karthik} 
\affiliation{International Centre for Theory of Quantum Technologies, University of Gd{\'a}nsk, Gd{\'a}nsk, Poland}
\email{karthik.hs@ug.edu.pl}

\author{B. N. Karthik}
\affiliation{Department of Physics, Bangalore University, Bangalore-560 056, India}
\email{karthikbnj@gmail.com }

\date{\today}
\begin{abstract} 

Extending the mathematical framework of Phys. Rev. A {\bf 102}, 052419 (2020) we construct Lorentz invariant quantities of pure three-qubit states. This method serves as a bridge between the well-known local unitary (LU) invariants viz. concurrences and three-tangle of an arbitrary  three-qubit pure state  and the Lorentz invariants of its reduced two-qubit systems. 
 \end{abstract}
 \pacs{03.65.Ta, 03.67.Mn}

\maketitle
\section{Introduction} 

The use of entanglement as a resource in quantum information processing tasks has accelerated research efforts on  its quantification, characterization, and control over the past two decades~\cite{amico2008,horodecki2009, stewart2014,chitambar2019,eberly2021}.  While multipartite entanglement poses higher level of complexity than the bipartite case, it enriches our theoretical understanding and paves way for innovative applications in distributed quantum networks~\cite{kempe1999,brub2017, huber2018, cunha2019,dur2020, wolfe2020, brub2022}. It has been recognized that geometry associated with particular symmetry transformations plays a vital role in exploring  multipartite entanglement --  especially in the distribution of entanglement  among the constituent subsystems~\cite{mahler1996,dur2000, ckw2000, acin2000, sudbery2001, brun2001, rau2021}.  Study of geometric invariants and canonical forms of composite quantum states  under {\em local} symmetry operations on subsystsems serves as a powerful tool to probe different manifestations of entanglement. To this end, considerable progress has been evinced in exploring local invariant quantities, canonical forms of equivalence classes of states under local unitary (LU) transformations, stochastic local operations and classical communication (SLOCC)~\cite{grassl1998, linden1998, fei2000, jaeger2001, ver2001, ver2002, preeti2003,  patricot2003, osterloh2010,wootters2011, kraus2012, tajima2013,torun2014,jens2015,meyer2017,sun2017,li2018,torun2019,supra2020,anjali2022,divya2023}. 

An essential feature of entanglement is that it remains  invariant under LU operations. Any two arbitrary pure states $\vert\psi\rangle$ and $\vert\phi\rangle$ are LU equivalent  (written symbolically as $\vert\psi\rangle~\sim~\vert\phi\rangle$) if and only if  they can be transformed into each other by local unitary operations.  A complete set of polynomial quantities that remain unaltered under LU operations on subsystems are used to certify  LU equivalence of multipartite states.  Recognizing normal/canonical form of a composite system by using LU transformations on individual subsystems is advantageous in evaluating these polynomial invariants. 

Ac{\'i}n et al. showed that  a three-qubit pure state under LU transformations can be reduced to a  canonical form given by~\cite{acin2000}:
\begin{equation}
	\label{3can}
	\vert\psi_{ABC}\rangle=\lambda_0\vert 0,0,0\rangle+\lambda_1 e^{i\phi}\vert 1,0,0\rangle+\lambda_2\vert 
	1,0,1\rangle+\lambda_3\vert 
	1,1,0\rangle+\lambda_4\vert 1,1,1\rangle
\end{equation} 
in terms of {\em five}  real entanglement parameters $\lambda_i\geq 0,\ i=0,1,2,3,4$ satisfying $\sum_{i=0}^4\,\lambda_i^2=1$, and a phase $\phi$ ranging between 0 and $\pi$. This gives a minimal form of pure three-qubit states containing only five terms and is helpful for evaluating LU invariants. 

We consider a  set of {\em five} LU invariants~\cite{acin2000} characterizing  pure three-qubit states (apart from normalization): 
\begin{eqnarray}
	\label{sudberyI}
	I_1&=&{\rm Tr}\, [\rho_{ BC}^2]={\rm Tr}\, [\rho_{ A}^2]= 1-2\lambda_0^2(1-\lambda_0^2-\lambda_1^2),\nonumber \\
	I_2&=&{\rm Tr}\, [\rho_{AC}^2]={\rm Tr} [\rho_{ B}^2] = 1-2\lambda_0^2(1-\lambda_0^2-\lambda_1^2-\lambda_2^2)-2\bigtriangleup , \\
	I_3&=&{\rm Tr}\, [\rho_{AB}^2]={\rm Tr}\,[\rho_{ C}^2] =1-2\lambda_0^2(1-\lambda_0^2-\lambda_1^2-\lambda_3^2)-2\bigtriangleup , \nonumber \\
	I_4&=&  {\rm Tr}\,[(\rho_{ A}\otimes\rho_{ B})\,\rho_{AB}]= 1+\lambda_0^2\,\left(\lambda_2^2 \lambda_3^2-\lambda_1^2 \lambda_4^2-2\,\lambda_2^2 -3\,\lambda_3^2-3\,\lambda_4^2\,\right)-(2-\lambda_0^2)\,\bigtriangleup  \nonumber\\
	I_5&=& \lambda_0^4\lambda_4^4=\frac{\tau^2}{16}, \nonumber 
\end{eqnarray} 
where 
 $\rho_A={\rm Tr}_{\rm B,C}\, \vert\psi_{\rm ABC}\rangle\langle \psi_{\rm ABC}\vert$, $\rho_B={\rm Tr}_{\rm A,C}\, \vert\psi_{\rm ABC}\rangle\langle \psi_{\rm ABC}\vert$, $\rho_C={\rm Tr}_{\rm A,B}\, \vert\psi_{\rm ABC}\rangle\langle \psi_{\rm ABC}\vert$, $\rho_{AB}~=~{\rm Tr}_{\rm C}\, \vert\psi_{\rm ABC}\rangle\langle \psi_{\rm ABC}\vert$, $\rho_{BC}={\rm Tr}_{\rm A}\, \vert\psi_{\rm ABC}\rangle\langle \psi_{\rm ABC}\vert$, $\rho_{AC}={\rm Tr}_{\rm B}\, \vert\psi_{\rm ABC}\rangle\langle \psi_{\rm ABC}\vert$ and 
\begin{equation}
\label{delta}
\bigtriangleup\equiv\vert \lambda_1\lambda_4e^{i\phi}-\lambda_2\lambda_3\vert^2. 
\end{equation} 
The first three invariants $I_1,\, I_2,\, I_3$  are related to the squares of the three one-to-other bipartite concurrences $C^2_{A(BC)},\ C^2_{B(AC)},$ and $C^2_{C(AB)}$ respectively~\cite{wootters1998,ckw2000}.  The fourth one $I_4$, is related to the {\em Kempe invariant}~\cite{kempe1999,sudbery2001} 
\begin{eqnarray}
	\label{i4}
{\cal I}_4&=& 3{\rm Tr}\,[(\rho_{ A}\otimes\rho_{ B})\,\rho_{AB}]-{\rm Tr}\,[\rho_{A}^3]-{\rm Tr}\,[\rho_{B}^3], \nonumber \\ 
& =& 3{\rm Tr}\,[(\rho_{ B}\otimes\rho_{ C})\,\rho_{BC}]-{\rm Tr}\,[\rho_{B}^3]-{\rm Tr}\,[\rho_{C}^3], \\ 
& =& 3{\rm Tr}\,[(\rho_{ A}\otimes\rho_{ C})\,\rho_{AC}]-{\rm Tr}\,[\rho_{A}^3]-{\rm Tr}\,[\rho_{C}^3], \nonumber 
\end{eqnarray}
which is symmetric under the permutation of qubits.  This quantity, while algebraically independent of the other LU invariants,  has no known implication towards the classification of three-qubit entanglement~\cite{osterloh2010}. 

Writing $\vert\psi_{\rm ABC}\rangle~=~\sum_{i,j,k=0,1}\, c_{ijk}\,\vert i,j,k\rangle$ in the computational basis, the invariant $I_5$ (Cayley’s hyperdeterminant~\cite{gelfand1994, miyake2003})   is expressed as  
\begin{equation}
	\label{i5}
I_5=\frac{1}{4}\,\left\vert\, \epsilon_{i_1\,i_2}\,\epsilon_{i_3\,i_4}\,\epsilon_{j_1\,j_2}\,\epsilon_{j_3\,j_4}\, \epsilon_{k_1\,k_3}\,\epsilon_{k_2\,k_4}\, c_{i_1j_1k_1}c_{i_2j_2k_2}c_{i_3j_3k_3}c_{i_4j_4k_4}\right\vert^2,
\end{equation}
where $\epsilon_{ij}$ denote antisymmetric tensor of rank-2; repeated indices are to be summed over in (\ref{i5}).  In Ac{\'i}n's canonical form (\ref{3can}) of the three-qubit state one obtains a  simple form  $I_5~=~\lambda_0^4\lambda_4^4$, which is related to the three-tangle $\tau=4\,\lambda_0^2\lambda_4^2$, a measure of three-way entanglement of three qubits in a pure state ~\cite{ckw2000}.

Any two pure three-qubit states $\vert\psi\rangle$ and $\vert\phi\rangle$ are  SLOCC equivalent if and only if they are mutually interconvertible  by means of local invertible transformations:  
\begin{equation}
	\vert\psi\rangle \sim A\otimes B\otimes C\, \vert\phi\rangle
\end{equation}
where  $A,\,B, \, C\in\ $SL(2,C)  denote $2\times 2$ complex  matrices with determinant unity.  Because  local protocols are unable to generate entanglement, invariant quantities under SLOCC are used for classification and also quantification of entanglement.  Equivalence classes  of pure three-qubit states under SLOCC was  explored in the celebrated work by  D{\"u}r  et al.~\cite{dur2000}, where it was shown that there exist {\em two} inequivalant  tripartite entanglement classes under SLOCC --  represented  by  the Greenberger-Horne-Zeilinger (GHZ) state~\cite{ghz1989}
\begin{equation}
	\label{ghz}
	\vert {\rm GHZ}\rangle =\frac{1}{\sqrt{2}}(\vert 0,0,0\rangle +\vert 1,1,1\rangle) 
\end{equation}
and the W state~\cite{dur2000} 
\begin{equation}
	\label{w}
	\vert {\rm W}\rangle =\frac{1}{\sqrt{3}}(\vert 1,0,0\rangle+\vert 0,1,0\rangle +\vert 0,0,1\rangle).
\end{equation}
 There has been a large  effort towards gaining deeper insight into the structure of SLOCC invariant quantities, 
 where three-qubit pure state is considered as a test bed~\cite{ckw2000,ver2002,patricot2003, osterloh2010, wootters2011,kraus2012,tajima2013,jens2015,li2018}.   
 
In this paper we extend the mathematical framework of Ref.~\cite{supra2020}  to construct Lorentz invariants of pure three-qubit states.  In the following section we describe the basic formalism   of Ref.~\cite{supra2020}. Mainly we highlight here that  the SLOCC property of the real $4\times 4$ matrix parametrization $\Lambda_{AB}$ of a two-qubit density matrix $\rho_{AB}$ paves the way to identify {\em Lorentz invariance} of the  eigenvalues  $\mu^{AB}_\alpha,\alpha=0,1,2,3$ of the matrix $\Gamma_{AB}=G\,\Lambda_{AB}\,G\,\Lambda^T_{AB}$.  Section 3 is devoted to explore the properties of the Lorentz invariant eigenvalues of $\Gamma_{ij}, \ ij=AB,\ BC,\ AC$ associated with the reduced two-qubit density matrices $\rho_{ij}$  of a pure three qubit state. We recognize that (i) the matrices $\Gamma_{ij}, \ ij=AB,\ BC,\ AC$ associated with a pure three qubit state have at most {\em two} distinct Lorentz invariant eigenvalues; (ii) difference between the two eigenvalues is symmetric under the interchange of qubits and is equal to the three-tangle $\tau$ of the three-qubit state; (iii) the smallest Lorentz invariant eigenvalue of $\Gamma_{ij}$ is equal to the squared concurrence $C^{2}_{ij},  \ ij=AB,\ BC,\ AC$ of the two-qubit subsystems. We  illustrate these features in pure permutation symmetric three-qubit states in subsection 3A.  Construction of a set of {\em five} SLOCC invariants,   which turn out to be the algebraic analogues of corresponding set of LU invariants of the three-qubit pure state, is outlined in subsection 3B.  A summary of our results is given in  Sec.4. 

 \section{Transformation of two-qubit state under SLOCC}
 Let us consider an arbitrary two-qubit density matrix $\rho_{AB}$, expanded in the Hilbert-Schmidt basis $\{\sigma_\alpha\otimes \sigma_\beta, \alpha,\beta=0,1,2,3\}$: 
 \begin{eqnarray}
 	\label{rho2q}
 	\rho_{AB}&=&\frac{1}{4}\, \sum_{\alpha,\,\beta=0}^{3}\,   
 	\left(\Lambda_{AB}\right)_{\alpha \, \beta}\, \left( \sigma_\alpha\otimes\sigma_\beta \right), \nonumber \\ 
 	\label{lambda}
 &&\left(\Lambda_{AB}\right)_{\alpha \, \beta}= {\rm Tr}\,\left[\rho_{AB}\,
 	(\sigma_\alpha\otimes\sigma_\beta)\,\right]  
 \end{eqnarray}
 where
 \begin{eqnarray}
 	\label{sigmamu}
 	\sigma_0=\left(\begin{array}{cc} 1 & 0 \\ 0 & 1       \end{array}\right),\ \  \sigma_1=\left(\begin{array}{cc} 0 & 1 \\ 1 & 0       \end{array}\right),\ \   \sigma_2=\left(\begin{array}{cc} 0 & -i \\ i & 0       \end{array}\right),\ \   \sigma_3=\left(\begin{array}{cc} 1 & 0 \\ 0 & -1       \end{array}\right).   
 \end{eqnarray} 
It is convenient to express the expansion coefficients $\left(\Lambda_{AB}\right)_{\mu \, \nu}$ in (\ref{rho2q}) as a $4\times 4$ real matrix: 
 \begin{eqnarray}
 	\label{lab}
 	\Lambda_{AB}&=&\left(\begin{array}{llll} 1& s_{B1} & s_{B2} & s_{B3} \\ 
 		s_{A1} & t^{AB}_{11}  & t^{AB}_{12} & t^{AB}_{13} \\ 
 		s_{A2} & t^{AB}_{21}  & t^{AB}_{22} & t^{AB}_{23} \\
 		s_{A3} & t^{AB}_{31}  & t^{AB}_{32} & t^{AB}_{33} \\
 	\end{array}\right).  
 \end{eqnarray} 
 Here $\mathbf{s}_{A}$, $\mathbf{s}_{B}$ are Minkowski four-vectors with components  $s_{A\alpha}$,  $s_{B\alpha},\ \alpha=0,1,2,3$ respectively and $T^{AB}=(t^{AB}_{ij}), \ i,\,i=1,2,3$ denotes the two-qubit correlation matrix: 
 \begin{eqnarray} 
 	\label{ri}
 	s_{A\alpha}&=& \left(\Lambda_{AB}\right)_{\alpha \, 0}={\rm Tr}\,\left[\rho_{AB}\, (\sigma_\alpha\otimes\sigma_0)\,\right]={\rm Tr}\,\left[\rho_{A}\,\sigma_\alpha\right], \\
 	\label{sj} 
 	s_{B\alpha}&=&   \left(\Lambda_{AB}\right)_{0 \, \alpha}= {\rm Tr}\,\left[\rho_{AB}\, (\sigma_0\otimes\sigma_\alpha)\,\right]={\rm Tr}\,\left[\rho_{B}\,\sigma_\alpha\right], \ \alpha=0,1,2,3  \\ 
 	\label{tij}
 	t^{AB}_{ij}&=& \left(\Lambda_{AB}\right)_{i \, j}= {\rm Tr}\,\left[\rho_{AB}\, (\sigma_i\otimes\sigma_j)\,\right],\  \ \ \ i,\,j=1,\,2,\,3. 
 \end{eqnarray} 
Under SLOCC the two-qubit state $\rho_{AB}$ transforms as    
 	\begin{eqnarray}
 		\rho_{AB}\longrightarrow\bar{\rho}_{AB}&=&\frac{(A\otimes B)\, \rho_{AB}\, (A^\dag\otimes B^\dag)}
 		{{\rm Tr}\left[\rho_{AB}\, (A^\dag\, A\otimes B^\dag\, B)\right]}
 	\end{eqnarray} 
 	where $A, B\in {\rm SL(2,C)}$ denote $2\times 2$ complex matrices with unit determinant. As a result one finds that    
 	\begin{eqnarray}
 		\label{sl2c}
 		\Lambda_{AB}\longrightarrow \bar{\Lambda}_{AB}&=&\frac{L_A\,\Lambda_{AB}\, L^T_B}{\left(L_A\,\Lambda_{AB}\, L^T_B\right)_{00}}
 	\end{eqnarray} 
 	where $L_A,\, L_B\in SO(3,1)$ are $4\times 4$  proper orthochronous Lorentz transformation matrices~\cite{kns1998} corresponding to  $A$, 
 	$B\in {\rm SL(2,C)}$ respectively and the superscript `$T$' denotes transpose operation.  
 
 We construct a $4\times 4$ real  matrix 
 \begin{equation}
 	\label{odef}
 	\Gamma_{AB}=G\,\Lambda_{AB}\, G\, \Lambda_{AB}^T,
 \end{equation}
 where $G={\rm diag}\,(1,-1,-1,-1)$ denotes the Minkowski metric~\cite{kns1998}. It is readily identified that the matrix $\Gamma_{AB}$ 	undergoes a similarity transformation under SLOCC (upto an overall  factor)~\cite{supra2020}:   
 	\begin{eqnarray}
 		\label{oa} 
 		\Gamma_{AB}\rightarrow \bar{\Gamma}_{AB}&=& G\,\bar{\Lambda}_{AB}\, G\, \bar{\Lambda}_{AB}^{T} \nonumber \\
 		&=& G\,L_{A}\, \Lambda_{AB}\, L_{B}^T\, G \, L_{B}\, \Lambda_{AB}^T L_{A}^T \nonumber \\ 
 		&=& \left(G\,L_{A}\,G\right)\, G\,\Lambda_{AB}\, \left(L_{B}^T\, G \, L_{B}\right)\, \Lambda_{AB}^T L_{A}^T \nonumber \\ 
 		&=& \left(L_{A}^T\right)^{-1}\, \Gamma_{AB}\, L_{A}^T  
 	\end{eqnarray}
 	where the defining property~\cite{kns1998}   $L^T\,G\,L=G$   of Lorentz transformation is used.  

 The matrix $\Gamma_{AB}$, constructed using the real matrix parametrization  $\Lambda_{AB}$ of the two-qubit density matrix $\rho_{AB}$ (see (\ref{lambda}), (\ref{lab})) exhibits the following important properties (see Theorem of Ref.~\cite{supra2020} on the nature of eigenvalues and eigenvectors of the
 matrix $\Gamma_{AB}$):
 		\begin{itemize}
 		\item[(i)] It  possesses  {\em non-negative} eigenvalues $\mu^{AB}_0\geq\mu^{AB}_1\geq\mu^{AB}_2\geq \mu^{AB}_3\geq 0$.  
 		\item [(ii)]   Four-eigenvector $X$ associated with the highest eigenvalue $\mu^{AB}_0$ of  the matrix $\Gamma_{AB}$ satisfies one of the following Lorentz invariant properties: 
 		\begin{equation}
 			\label{positive}
 	X^T\, G\, X>0   \\
 		\end{equation}
 		or
 		\begin{equation}
 			\label{neutral} 
 			X^T\, G\, X =0.  
 		\end{equation} 
	The condition (\ref{neutral}) is accompanied by the observation that the matrix $\Gamma_{AB}$ has only {\em two}  eigenvalues $\mu^{AB}_0$, $\mu^{AB}_2$ with $ \mu^{AB}_0\geq \mu^{AB}_2$, both of which are  doubly degenerate.
 		\item [(iii)] Suppose the eigenvector  $X$ satisfies the Lorentz invariant condition (\ref{positive}). Then there exists  suitable SLOCC transformations $A_1,\, B_1\in {\rm SL(2,C)}$ (with corresponding    Lorentz transformations $L_{A_1},\, L_{B_1}\in SO(3,1)$ respectively) such that the matrix $\Gamma_{AB}$ assumes a diagonal canonicial form: 
 		\begin{eqnarray} 
 			\label{o1c}
 			\bar{\Gamma}_{AB}^{(I_c)}&=& \left(L^T_{A_{I_c}}\right)^{-1}\, \Gamma_{AB}\, L^T_{A_{I_c}}  ={\rm diag}  \,  \left(\mu^{AB}_0,\, \mu^{AB}_1,\, \mu^{AB}_2,\, \mu^{AB}_3\right). 
 		\end{eqnarray}
 	 \item[(iv)] Associated with the standard form $\bar{\Gamma}_{AB}^{(I_c)}$ it is seen that~\cite{supra2020}  
 	\begin{eqnarray*}	 
 		\rho_{AB}\longrightarrow\bar{\rho}^{\,I_c}_{AB}&=&\frac{(A_{I_c}\otimes B_{I_c})\, \rho_{AB}\, (A_{I_c}^\dag\otimes B_{I_c}^\dag)}
 		{{\rm Tr}\left[\rho_{AB}\, (A_{I_c}^\dag\, A_{I_c} \otimes B_{I_c}^\dag\, B_{I_c})\right]}
 	\end{eqnarray*} 
 	reduces to the Bell-diagonal form  
 	\begin{eqnarray}	 
 		\label{rhobd} 
 		\bar{\rho}^{\,I_c}_{AB}
 		&=&  \frac{1}{4}\, \left( \sigma_0\otimes \sigma_0 + \sum_{i=1,2}\, \sqrt{\frac{\mu^{AB}_i}{\mu^{AB}_0}}\, \sigma_i\otimes\sigma_i \pm \sqrt{\frac{\mu^{AB}_3}{\mu^{AB}_0}}\, \sigma_3\otimes\sigma_3 \right)
 	\end{eqnarray}
 	under SLOCC. Here the sign $\pm$ is chosen based on  sgn$[\det(\Lambda_{AB})]=\pm$. 
 	\item [(v)] Whenever the eigenvector  $X$ obeys the condition (\ref{neutral})  suitable SLOCC transformations 
 	$A_{II_c},\, B_{II_c}\in SL(2,C)$ (associated  Lorentz transformations denoted respectively by $L_{A_{II_c}},\, L_{B_{II_c}}\in SO(3,1)$) exist such that the real symmetric matrix $\Gamma_{AB}$ takes a non-diagonal canonical form:  
 	\begin{eqnarray}
 		\label{o2c}
 		\bar{\Gamma}_{AB}^{(II_c)}&=& L_{A_{II_c}}\, \Gamma_{AB}\, L^T_{A_{II_c}} =      
 		\begin{pmatrix}
 			\phi^{AB}_0 & 0  & 0 & \phi^{AB}_0-\mu^{AB}_0 \\ 
 			0 & \mu^{AB}_2 & 0 & 0 \\ 
 			0 & 0 &   \mu^{AB}_2 & 0 \\
 			\mu_0^{AB}-\phi^{AB}_0 & 0 & 0 &  2\,\mu^{AB}_0-\phi^{AB}_0, 
 		\end{pmatrix}        
 	\end{eqnarray}
 	where 
 	\begin{equation}
 		\phi^{AB}_0=\left(L_{A_{II_c}}\,\Gamma_{AB}\,L_{A_{II_c}}^T\right)_{00}.
 	\end{equation}
\item[(vi)] Consequently, the canonical form of the two-qubit density matrix is given by 
\begin{eqnarray}
	\label{rho2c} 
	\rho_{AB}\longrightarrow\bar{\rho}^{\,II_c}_{AB}&=&\frac{(A_{II_c}\otimes B_{II_c})\, \rho_{AB}\, (A_{II_c}^\dag\otimes B_{II_c}^\dag)}
	{{\rm Tr}\left[\rho_{AB}\, (A_{II_c}^\dag\, A_{II_c} \otimes B_{II_c}^\dag\, B_{II_c})\right]},  \\ 
	&=&  \frac{1}{4}\, \left[ \, \sigma_0\otimes \sigma_0 + (1-\gamma^{AB}_0)\,\sigma_3\otimes \sigma_0 + \gamma^{AB}_2 \,(\sigma_1\otimes\sigma_1 - \sigma_2\otimes\sigma_2) +\,\gamma^{AB}_0\, \sigma_3\otimes\sigma_3\right]  \nonumber  
\end{eqnarray}
where 
\begin{equation}
	\label{adef}
	\gamma^{AB}_0=\frac{\mu^{AB}_0}{\phi^{AB}_0},\ \ \ \ \  \gamma^{AB}_2=\sqrt{\frac{\mu^{AB}_2}{\phi^{AB}_0}}, \ \  \ 0\leq \left(\gamma^{AB}_2\right)^2\leq \gamma^{AB}_0\leq 1. 
\end{equation}
\item[(vii)] The eigenvalues $\mu^{AB}_\alpha,\ \alpha=0,1,2,3$ of the $4\times 4$ matrix $\Gamma_{AB}$ are Lorentz invariant i.e., they are unchanged under SLOCC.
 \end{itemize}

In the following section, we construct SLOCC invariants, by exploiting the Lorentz transformation properties of the real matrices  $\Lambda_{AB},\ \Lambda_{BC}$ and $\Lambda_{AC}$ characterizing the two-qubit subsystems of a pure three-qubit state. 

\section{Lorentz invariants of  pure three-qubit  state} 

Let us write the two-qubit reduced density matrices $\rho_{AB}$, $\rho_{BC}$ and $\rho_{AC}$ of a pure three-qubit state as   
\begin{eqnarray}
	\label{rab}
	\rho_{AB}&=&{\rm Tr}_{\rm C}\, \vert\psi_{\rm ABC}\rangle\langle \psi_{\rm ABC}\vert= \frac{1}{4}\, \sum_{\alpha,\,\beta=0}^{3}\, \left(\Lambda_{AB}\right)_{\alpha \, \beta}\, \left( \sigma_\alpha\otimes\sigma_\beta \right), \\ 
	\label{rbc}
	\rho_{BC}&=&{\rm Tr}_{\rm A}\, \vert\psi_{\rm ABC}\rangle\langle \psi_{\rm ABC}\vert= \frac{1}{4}\, \sum_{\alpha,\,\beta=0}^{3}\, \left(\Lambda_{BC}\right)_{\alpha \, \beta}\, \left( \sigma_\alpha\otimes\sigma_\beta \right), \\ 
	\label{rac}
	\rho_{AC}&=&{\rm Tr}_{\rm B}\, \vert\psi_{\rm ABC}\rangle\langle \psi_{\rm ABC}\vert= \frac{1}{4}\, \sum_{\alpha,\,\beta=0}^{3}\, \left(\Lambda_{AC}\right)_{\alpha \, \beta}\, \left( \sigma_\alpha\otimes\sigma_\beta \right). 
\end{eqnarray}
We employ Ac{\'i}n's canonical form (\ref{3can}) of the pure  three-qubit state, for evaluating  the $4\times 4$ real matrices   $\Lambda_{AB},\ \Lambda_{BC}$ and  $\Lambda_{AC}$ explicitly:  
{\scriptsize\begin{eqnarray}
	\label{lab3}
\Lambda_{AB}&=&\left(\begin{array}{cccc}   
1& 2\,(\lambda_2\,\lambda_4+\lambda_1\lambda_3\,\cos\phi) &  -2\, \lambda_1\,\lambda_3\,\sin\phi & 1-2\,(\lambda^2_3+\lambda^2_4) \\ 
2\,\lambda_0\,\lambda_1\,\cos\phi & 2\, \lambda_0\,\lambda_3 & 0 & 2\,\lambda_0\,\lambda_1\,\cos\phi \\ 
2\,\lambda_0\,\lambda_1\,\sin\phi & 0 &  -2\, \lambda_0\,\lambda_3 & 2\,\lambda_0\,\lambda_1\,\sin\phi \\ 
2\, \lambda_0^2-1 & -2\,(\lambda_2\,\lambda_4+\lambda_1\lambda_3\,\cos\phi) & 2\,\lambda_1\,\lambda_3\,\sin\phi & 1-2\,(\lambda^2_1+\lambda^2_2)
\end{array}\right), \\
\label{lbc}
\Lambda_{BC}&=&\left(\begin{array}{cccc}   
	1& 2\,(\lambda_3\,\lambda_4+\lambda_1\lambda_2\,\cos\phi) &  -2\, \lambda_1\,\lambda_2\, \sin\phi\, & 1-2\,(\lambda^2_2+\lambda^2_4) \\ 
	2\,(\lambda_2\,\lambda_4+\lambda_1\lambda_3\,\cos\phi) & 2\,(\lambda_2\,\lambda_3+\lambda_1\lambda_4\,\cos\phi) & -2\, \lambda_1\,\lambda_4\, \sin\phi  & -2\,(\lambda_2\,\lambda_4-\lambda_1\lambda_3\,\cos\phi) \\ 
-2\, \lambda_1\,\lambda_3\, \sin\phi & -2\, \lambda_1\,\lambda_4\, \sin\phi &  2\,(\lambda_2\,\lambda_3-\lambda_1\lambda_4\,\cos\phi) & -2\,\lambda_1\,\lambda_3\,\sin\phi \\ 
	1-2\,(\lambda^2_3+\lambda^2_4) & -2\,(\lambda_3\,\lambda_4-\lambda_1\lambda_2\,\cos\phi) & -2\,\lambda_1\,\lambda_2\,\sin\phi & 1-2\,(\lambda^2_2+\lambda^2_3)
\end{array}\right),\\
\Lambda_{AC}&=&\left(\begin{array}{cccc}   
	1& 2\,(\lambda_3\,\lambda_4+\lambda_1\lambda_2\,\cos\phi) &  -2\, \lambda_1\,\lambda_2\,\sin\phi & 1-2\,(\lambda^2_2+\lambda^2_4) \\ 
	2\,\lambda_0\,\lambda_1\,\cos\phi & 2\, \lambda_0\,\lambda_2 & 0 & 	2\,\lambda_0\,\lambda_1\,\cos\phi \\ 
	2\,\lambda_0\,\lambda_1\,\sin\phi & 0 &  -2\, \lambda_0\,\lambda_2 & 2\,\lambda_0\,\lambda_1\,\sin\phi \\ 
	2\, \lambda_0^2-1 & -2\,(\lambda_3\,\lambda_4+\lambda_1\lambda_2\,\cos\phi) & 2\,\lambda_1\,\lambda_2\,\sin\phi & 1-2\,(\lambda^2_1+\lambda^2_3)
\end{array}\right).
\end{eqnarray}}

Let us recall the formula for the concurrence $C_{AB}$ of an arbitrary two-qubit state $\rho_{AB}$ introduced by Wootters~\cite{wootters1998}: 
\begin{equation}
	C_{AB}={\rm max}\,\{0,\nu^{AB}_1-\nu^{AB}_2-\nu^{AB}_3-\nu^{AB}_4\}
\end{equation} 
where $\nu^{AB}_i,\ i=1,2,3,4$ are the square roots of the eigenvalues of 
\begin{equation}
	\rho_{AB}\,\widetilde{\rho}_{AB}= \rho_{AB}\,(\sigma_2\otimes\sigma_2)\, \rho_{AB}^T\, (\sigma_2\otimes\sigma_2)
\end{equation}
in decreasing order. While the matrix $\rho_{AB}\,\widetilde{\rho}_{AB}$ is non-hermitian, it has only real and positive eigenvalues~\cite{wootters1998}. 

To gain further insight into the structure of the non-hermitian matrix $\rho_{AB}\,\widetilde{\rho}_{AB}$, we express the spin flipped two-qubit density matrix $\widetilde{\rho}_{AB}=(\sigma_2\otimes\sigma_2)\, \rho_{AB}^T\, (\sigma_2\otimes\sigma_2)$ in the basis $\{\sigma_\alpha\otimes \sigma_\beta, \alpha,\beta=0,1,2,3\}$ to obtain
\begin{equation}
	\label{trho}
	\widetilde{\rho}_{AB}= \frac{1}{4}\, \sum_{\alpha,\beta=0}^3\, \left(\Lambda'_{AB}\right)_{\alpha\,\beta}\, \sigma_\alpha\otimes \sigma_\beta
\end{equation}
where the $4\times 4$ real matrix $\Lambda'_{AB}$ characterizing the spin flipped two-qubit density matrix  $\widetilde{\rho}_{AB}$ is found to be~\cite{fn1} 
\begin{equation}
	\label{lprime}
	\Lambda'_{AB}=G\,\Lambda_{AB}\,G. 
\end{equation}
We thus recognize (see (\ref{rho2q}),(\ref{trho}), (\ref{lprime}) and (\ref{odef})) that 
\begin{eqnarray}
	\label{trab}
	{\rm Tr}[\rho_{AB}\,\widetilde{\rho}_{AB}]= \frac{1}{4}\, {\rm Tr}\,[G\,\Lambda_{AB}\,G\,\Lambda_{AB}^T]=\frac{1}{4}\, {\rm Tr}\,[\Gamma_{AB}].
\end{eqnarray}
Evidently,  ${\rm Tr}\,[\Gamma_{AB}]\geq 0$ as the $4\times 4$ real matrix $\Gamma_{AB}=G\,\Lambda_{AB}\,G\,\Lambda_{AB}^T$ is  non-negative~\cite{supra2020} and this, in turn, justifies that  trace of the non-hermitian matrix  $\rho_{AB}\,\widetilde{\rho}_{AB}$ (LHS of (\ref{trab})) is also positive.

In a pure entangled three-qubit state  every pair of qubits are entangled with the remaining qubit. Thus, the two-qubit subsystem density matrix of a pure three-qubit state has at most \break {\em two}
nonzero eigenvalues. As a result,  the matrix $\rho_{ij}\,\widetilde{\rho}_{ij}$ has only {\em two} non-zero eigenvalues $\left(\nu^{ij}_1\right)^2,\left(\nu^{ij}_2\right)^2,\ ij=AB, BC,AC$. Thus, the squared concurrence $C^2_{ij}$ of two-qubit subsystem state $\rho_{ij}$ of a  three-qubit pure state simplifies to
\begin{equation}
	\label{con2}
	C^2_{ij}=(\nu^{ij}_1-\nu^{ij}_2)^2, \ \ ij=AB, BC,AC. 
\end{equation}

We are interested in recognizing SLOCC invariants of three-qubit pure state, which are useful in determining the Lorentz invariant eigenvalues of the matrices $\Gamma_{AB}, \ \Gamma_{BC}$ and $\Gamma_{AC}.$ 
In Ac{\'i}n's canonical form (\ref{3can}) the  matrices $\Gamma_{AB}, \ \Gamma_{BC}$ and $\Gamma_{AC}$ have the following explicit structure: 
\begin{eqnarray}
	\label{gabe} 
\Gamma_{AB}&=&\left(\begin{array}{cccc}   
	4\, \lambda_0^2 (\lambda_2^2+\lambda_3^2+\lambda_4^2) & 	4\, \lambda_0^2\, \lambda_2\, \lambda_4  &  0 & 	4\, \lambda_0^2\, \lambda_2^2 \\ 
	-4\, \lambda_0^2\, \lambda_2\, \lambda_4 & 4\, \lambda_0^2\, \lambda_3^2  & 0 & 	-4\, \lambda_0^2\, \lambda_2\, \lambda_4  \\ 
	0 & 0 &  4\, \lambda_0^2\, \lambda_3^2 & 0    \\ 
-4\, \lambda_0^2\, \lambda_2^2 & -4\, \lambda_0^2\, \lambda_2\, \lambda_4  & 0 & 4\, \lambda_0^2 (\lambda_3^2-\lambda_2^2+\lambda_4^2) 
\end{array}\right) \\ 
\label{gbce} 
\Gamma_{BC}&=&\left(\begin{array}{cccc}   
	4\, \lambda_0^2\,(\lambda_3^2+\lambda_4^2)+4\,\bigtriangleup) & 	4\, \lambda_0^2\, \lambda_3\, \lambda_4  &  0 & 	4\, \lambda_0^2\, \lambda_3^2 \\ 
	-4\, \lambda_0^2\, \lambda_3\, \lambda_4 & 4\, \bigtriangleup & 0 & 	-4\, \lambda_0^2\, \lambda_3\, \lambda_4  \\ 
	0 & 0 &  4\,\bigtriangleup & 0    \\ 
	-4\, \lambda_0^2\, \lambda_3^2 & -4\, \lambda_0^2\, \lambda_3\, \lambda_4  & 0 & 	4\, \lambda_0^2\,(\lambda_4^2-\lambda_3^2)+4\,\bigtriangleup)
\end{array}\right) \\ 
	\label{gace} 
\Gamma_{AC}&=&\left(\begin{array}{cccc}   
	4\, \lambda_0^2 (\lambda_2^2+\lambda_3^2+\lambda_4^2) & 	4\, \lambda_0^2\, \lambda_3\, \lambda_4  &  0 & 	4\, \lambda_0^2\, \lambda_3^2 \\ 
	-4\, \lambda_0^2\, \lambda_3\, \lambda_4 & 4\, \lambda_0^2\, \lambda_2^2  & 0 & 	-4\, \lambda_0^2\, \lambda_3\, \lambda_4  \\ 
	0 & 0 &  4\, \lambda_0^2\, \lambda_2^2 & 0    \\ 
	-4\, \lambda_0^2\, \lambda_3^2 & -4\, \lambda_0^2\, \lambda_3\, \lambda_4  & 0 & 4\, \lambda_0^2 (\lambda_2^2-\lambda_3^2+\lambda_4^2) 
\end{array}\right).   
\end{eqnarray} 
\begin{itemize}
\item We find that these matrices $\Gamma_{AB}$, $\Gamma_{BC}$ and $\Gamma_{AC}$ have {\em at most} two distinct eigenvalues:    
\begin{eqnarray} 
	\label{egab}
	\mu_0^{AB}&=&\mu_1^{AB}=4\, \lambda_0^2\,(\lambda_3^2+\lambda_4^2),\   	\mu_2^{AB}=\mu_3^{AB}=4\, \lambda_0^2\,\lambda_3^2, \\
	\label{egbc}
		\mu_0^{BC}&=&\mu_1^{BC}=4\, (\bigtriangleup + \lambda_0^2\,\lambda_4^2), \	\mu_2^{BC}==\mu_3^{BC}=4\, \bigtriangleup, \\
	\label{egac}
	\mu_0^{AC}&=&\mu_1^{AC}=4\, \lambda_0^2\,(\lambda_2^2+\lambda_4^2),\   	\mu_2^{AC}=\mu_3^{AC}=4\, \lambda_0^2\,\lambda_2^2.
\end{eqnarray}

\item We  notice an interesting feature that the differences between the largest and the smallest eigenvalues of $\Gamma_{AB}$, $\Gamma_{BC}$, $\Gamma_{AC}$ are  identically equal to the three-tangle $\tau$  of  the three-qubit state: 
\begin{equation}
	\label{diff}
\mu_0^{ij}-\mu_2^{ij}= 4\, \lambda_0^2\,\lambda_4^2=\tau, \ \ \ \ ij=AB, BC,AC. 
\end{equation}
This reveals the fact that the LU invariant $I_5=\frac{\tau^2}{16}$  of the three-qubit state (see last line of (\ref{sudberyI})) is a permutation symmetric  SLOCC invariant. It is worth noting that $\sqrt{I_5}=\lambda_0^2\,\lambda_4^2$ is equal to the the product\cite{ckw2000} $\nu^{AB}_1\,\nu^{AB}_2=\nu^{BC}_1\,\nu^{BC}_2=\nu^{AC}_1\,\nu^{AC}_2$ of the square root of the eigenvalues of   $\rho_{ij}\,\widetilde{\rho}_{ij},\ ij=AB, BC,AC$. Thus, 
\begin{equation}
	\label{lrrtdiff}
	\mu_0^{ij}-\mu_2^{ij}= 4\, \nu^{ij}_1\,\nu^{ij}_2.
\end{equation}

\item The Lorentz invariant eigenvalues of $\Gamma_{AB}$, $\Gamma_{BC}$ and $\Gamma_{AC}$ can be determined using  $I_5$ along with {\em three} more  SLOCC invariants given by, 
\begin{eqnarray}
\label{invabc}
{\cal K}_1&=&\frac{1}{4}\,{\rm Tr}[\, \Gamma_{AB}]=\frac{1}{2}\,(\mu_0^{AB}+\mu_2^{AB})\nonumber \\ 
{\cal K}_2&=&\frac{1}{4}\,{\rm Tr}[\, \Gamma_{BC}]=\frac{1}{2}\,(\mu_0^{BC}+\mu_2^{BC}) \\ 
{\cal K}_3&=&\frac{1}{4}\,{\rm Tr}[\, \Gamma_{AC}]= \frac{1}{2}\, (\mu_0^{AC}+\mu_2^{AC}). \nonumber 
\end{eqnarray} 

Substituting (\ref{trab}) in (\ref{invabc})  we obtain 
\begin{eqnarray}
	\label{invrabc}
{\cal K}_1&=&	 {\rm Tr}[\rho_{AB}\,\widetilde{\rho}_{AB}]=(\nu^{AB}_1)^2+(\nu^{AB}_2)^2, \nonumber \\
	{\cal K}_2&=& {\rm Tr}[\rho_{BC}\,\widetilde{\rho}_{BC}]= (\nu^{BC}_1)^2+(\nu^{BC}_2)^2, \\ 
	{\cal K}_3&=& {\rm Tr}[\rho_{AC}\,\widetilde{\rho}_{AC}]= (\nu^{AC}_1)^2+(\nu^{AC}_2)^2.  \nonumber
\end{eqnarray}
\end{itemize}

We proceed to prove the following theorem: 
\begin{thm}
The squared concurrence $C^2_{ij}$ of the two-qubit subsystem $\rho_{ij}$  of a pure three-qubit state is  equal to  the smallest Lorentz  invariant eigenvalue $\mu^{ij}_{2}$ of the $4\times 4$  matrix  $\Gamma_{ij}~=~G\,\Lambda_{ij}\,G\,\Lambda^T_{ij},\ \ \ ij=AB,BC, AC.$ 
\end{thm}
\begin{pf}
Using   (\ref{lrrtdiff}), (\ref{invabc}) and (\ref{invrabc})  we connect the eigenvalues of $\Gamma_{AB},\ \Gamma_{BC},\ \Gamma_{AC}$ with  those of $\rho_{AB}\,\widetilde{\rho}_{AB},\ \rho_{BC}\,\widetilde{\rho}_{BC},\ \rho_{AC}\,\widetilde{\rho}_{AC}$ respectively: 
\begin{eqnarray}
	4\, \nu^{ij}_1 \nu^{ij}_2&=& \mu_0^{ij}-\mu_2^{ij}, \nonumber \\
	\left[ \,(\nu^{ij}_1)^2+(\nu^{ij}_2)^2\right]&=&\frac{1}{2}\,(\mu_0^{ij}+\mu_2^{ij}), \ \ \ \ ij=AB, BC,AC. 
\end{eqnarray}
We thus obtain (see (\ref{con2}))
\begin{eqnarray}
\mu_2^{ij}=\left(\nu^{ij}_1-\nu^{ij}_2\right)^2=C^2_{ij},\ \ \  ij=AB, BC,AC. 
 \end{eqnarray}
\end{pf}
The above theorem provides an alternate method to evaluate concurrences of two-qubit subsystems $\rho_{ij}, \ ij=AB,BC,AC$ of a pure three-qubit state, in terms of the smallest Lorentz invariant eigenvalues of $\Gamma_{ij}$.
\begin{flushright} $\Box$\end{flushright}

\subsection{Permutation symmetric three-qubit pure states} 
It is well-known that permutation symmetric states offer conceptual clarity and computational simplicity in the anlaysis of local invariants~\cite{meyer2017,anjali2022,uma2006,ijmp2006,uma2007}.   In this subsection we illustrate the effectiveness of our framework to evaluate concurrence and tangle in pure three-qubit permutation symmetric states, where we make use of the explicit parametrization given by Meill and Meyer~\cite{meyer2017} recently.

Consider a one-parameter familty of three-qubit permutation symmetric state~\cite{ meyer2017,anjali2022}: 
\begin{equation}
	\label{psiW}
\vert\psi_{\rm sym}(\beta)\rangle = \frac{1}{\sqrt{2+\cos\beta}}\, \left( \sqrt{3}\,\cos\frac{\beta}{2}\,\left\vert 0_A\,0_B\, 0_C\,\right\rangle + \sin\frac{\beta}{2}\,\left\vert{\rm W}\,\right\rangle\right)
\end{equation} 
where $0<\beta\leq\,\pi$. We find that~\cite{anjali2022}
\begin{eqnarray}
\Gamma_{\rm sym}(\beta)&=&\Gamma_{AB}(\beta)=\Gamma(\beta)_{BC}=\Gamma_{AC}(\beta)\nonumber \\ 
&=&\left(\begin{array}{cccc} 
		2\,u(\beta) & 0 &0 & u(\beta) \\
		0& u(\beta) & 0  &  0 \\ 
		0 & 0 &  u(\beta) & 0 \\ 
		-u(\beta)  &  0 & 0 & 0\end{array} \right),\ \ \ \ u(\beta)=\left[\frac{1-\cos\beta}{3(2+\cos\beta)}\right]^2. 
\end{eqnarray}
The Lorentz invariant eigenvalues of  $\Gamma_{\rm sym}(\beta)$ are equal i.e., 
\begin{equation}
	\label{muW}
\mu_{0}(\beta)=\mu_{2}(\beta)=\left[\frac{1-\cos\beta}{3(2+\cos\beta)}\right]^2
\end{equation}
and hence the concurrence of $\rho_{\rm sym}(\beta)=\rho_{AB}(\beta)=\rho_{BC}(\beta)=\rho_{AC}(\beta)$ is given by  
\begin{equation}
C(\beta)=\frac{1-\cos\beta}{3(2+\cos\beta)},
\end{equation} 
in perfect agreement with the result given by Meill and Meyer~\cite{meyer2017}. Substitution of (\ref{muW}) in the LHS of (\ref{diff}) confirms that the three-tangle $\tau(\beta)$ for the state (\ref{psiW}) is zero. 

We proceed further with a three-parameter family of pure three-qubit permutation symmetric state~\cite{meyer2017} 
\begin{eqnarray}
	\label{psiGHZ}
\vert\psi_{\rm sym}(y,\beta,\phi)\rangle&=&N \left(\vert 0\,\rangle^{\otimes\,3}  + y\,  
e^{i\, \phi} \, \vert \beta\rangle^{\otimes\,3} \right),
 \end{eqnarray}
where $\vert \beta\rangle=\cos\,\frac{\beta}{2}\, \vert 0\rangle+\sin\,\frac{\beta}{2}\,\vert 1\rangle,\ 0<y\leq 1, \ 0\leq \phi \leq 2\pi,\ 0<\beta\leq \pi.$ We evaluate  the matrix $\Gamma_{\rm sym}(y,\beta,\phi)\equiv\Gamma_{AB}(y,\beta,\phi)=\Gamma_{BC}(y,\beta,\phi)=\Gamma_{AC}(y,\beta,\phi)$  in the state (\ref{psiGHZ}) (see Ref.~\cite{anjali2022} for details):
\begin{eqnarray}
	\label{gGHZ}
\Gamma_{\rm sym}(y,\beta,\phi)&=&
	{\cal B}(y,\beta,\phi)\ba{cccc} 3+\cos \beta & \sin\beta & 0 & 1+\cos \beta  \\ 
	-\sin \beta & (1+\cos \beta) & 0 & -\sin \beta  \\ 0 & 0 & (1+\cos \beta) & 0 \\ 
	-(1+\cos \beta) & -\sin\beta & 0 & (1-\cos \beta) \ea 
\end{eqnarray} 
where 
\begin{equation}
\label{calb}
{\cal B}(y,\beta,\phi)=\frac{y^2 (1-\cos \beta)^2}{2\left(1+y^2+2\,y\cos\,\phi 
	\cos^3 \frac{\beta}{2} \right)^{2}}.
\end{equation}
The Lorentz invariant eigenvalues of 	$\Gamma_{\rm sym}(y,\beta,\phi)$ are found to be  
\begin{equation}
	\label{muGHZ}
	\mu_{0}(y,\beta,\phi)= 2\,{\cal B}(y,\beta,\phi), \ 	\mu_{2}(y,\beta,\phi)={\cal B}(y,\beta,\phi)\,\,(1+\cos \beta). 
\end{equation}
The concurrence for the two-qubit subsystem density matrices $\rho_{\rm sym}(y,\beta,\phi)=\rho_{AB}(y,\beta,\phi)=\rho_{BC}(y,\beta,\phi)=\rho_{AC}(y,\beta,\phi)$ drawn from the three-qubit pure symmetric state (\ref{psiGHZ}) is thus given by, 
\begin{eqnarray}
C(y,\beta,\phi)&=&\sqrt{{\cal B}(y,\beta,\phi)\,\,(1+\cos \beta)}\nonumber \\ 
                 &=& \frac{2\,y\, \sin\beta\, \sin\frac{\beta}{2}  }{\left(1+y^2+2\,y\cos\,\phi 
                 	\cos^3 \frac{\beta}{2} \right)}
\end{eqnarray}
which matches exactly with the formula derived in Ref.~\cite{meyer2017}. 

Substituting (\ref{muGHZ}) in (\ref{diff}) we evaluate the three-tangle $\tau(y,\beta,\phi)$ in (\ref{psiGHZ}) to obtain  
 \begin{eqnarray}
 \tau(y,\beta,\phi)&=& \mu_0(y,\beta,\phi)-\mu_{2}(y,\beta,\phi) \nonumber \\
& =& {\cal B}(y,\beta,\phi)(1-\cos \beta) \nonumber \\ 
&=& \left(\frac{2\,y\,  \sin^3\frac{\beta}{2}  }{1+y^2+2\,y\cos\,\phi 
	\cos^3 \frac{\beta}{2}}\right)^2,
  \end{eqnarray}
in agreement with  the expression for $\tau^2$ given in Ref.~\cite{meyer2017} for the state (\ref{psiGHZ}).

\subsection{LU versus SLOCC invariants of pure three-qubit state}

Taking a closer look at  the set of {\em five} LU invariants (\ref{sudberyI}) of pure three-qubit states,  it is seen that  $I_1={\rm Tr}[\rho_{AB}^2]$, $I_2={\rm Tr}[\rho_{AC}^2]$, $I_3={\rm Tr}[\rho_{AB}^2]$  get replaced by  their SLOCC counterparts (see (\ref{invrabc}))
${\cal K}_1={\rm Tr}[\rho_{AB}\,\widetilde{\rho}_{AB}]$, ${\cal K}_2={\rm Tr}[\rho_{BC}\,\widetilde{\rho}_{BC}]$, ${\cal K}_3={\rm Tr}[\rho_{AC}\,\widetilde{\rho}_{AC}]$. It is seen that the LU invariant  $I_5=\tau^2/16$ enjoys a higher level of invariance,  by remaining unchanged when a three-qubit pure state undergoes SLOCC. Thus, we have {\em four} SLOCC invariants, which encode information about the entanglement content in the three-qubit pure state since it is possible to reconstruct concurrences $C_{AB},\ C_{BC},\ C_{AC}$ and three-tangle $\tau$ using them. 
On the other hand, the Kempe invariant ${\cal I}_4$ given by (\ref{i4}) (which is a permutation symmetric extension of the LU invariant $I_4$ listed in (\ref{sudberyI})) is known to be algebraically independent of concurrences and three-tangles~\cite{osterloh2010}. In order to complete the set of SLOCC invariants, we study the structure of $I_4$ with an intention to find its Lorentz invariant analogue. Using (\ref{rho2q}), (\ref{ri}), (\ref{sj}), we obtain
\begin{equation}
\label{extni4}
I_4={\rm Tr}\,[(\rho_{ A}\otimes\rho_{ B})\,\rho_{AB}]=\frac{1}{4}\, \left(\mathbf{s}_{A}^T\,\Lambda_{AB}\, \mathbf{s}_B\right). 
\end{equation}
We consider~\cite{fn2} 
\begin{equation}
	\label{extnk4}
	{\cal K}_4={\rm Tr}\,[(\rho_{ A}\otimes\rho_{ B})\,\widetilde{\rho}_{AB}]=\frac{1}{4}\, \left(\mathbf{s}_{A}^T\,G\,\Lambda_{AB}\,G\,\mathbf{s}_B\right). 
\end{equation}
to be the Lorentz invariant analogue replacing $I_4$. 
Thus, we have the following set of {\em five} SLOCC invariants 
\begin{eqnarray}
	{\cal K}_1&=& 2\, {\rm Tr}[\rho_{AB}\,\widetilde{\rho}_{AB}]= 4\,\lambda_0^2\,\lambda_3^2+2\, \lambda_0^2\lambda_4^2= C^2_{AB}+\frac{\tau}{2}, \nonumber \\ 
	{\cal K}_2&=& 2\, {\rm Tr}[\rho_{BC}\,\widetilde{\rho}_{BC}]=4\,\bigtriangleup +2\, \lambda_0^2\lambda_4^2=C^2_{BC}+\frac{\tau}{2}, \nonumber \\ 
	{\cal K}_3&=& 2\, {\rm Tr}[\rho_{AC}\,\widetilde{\rho}_{AC}]=4\,\lambda_0^2\,\lambda_2^2+2\,\lambda_0^2\,\lambda_4^2=C^2_{AC}+\frac{\tau}{2}, \\ 
	{\cal K}_4&=&{\rm Tr}\,[(\rho_{ A}\otimes\rho_{ B})\,\widetilde{\rho}_{AB}]= \lambda_0^2\,\left(\bigtriangleup +\lambda_2^2 \lambda_3^2-\lambda_1^2 \lambda_4^2+ \lambda_3^2 +\lambda_4^2\right), \nonumber \\ 
	{\cal K}_5&\equiv& I_5=  \lambda_0^4\,\lambda_4^4=\frac{\tau^2}{16}, \nonumber 
\end{eqnarray}
which are the algebraic counterparts of the LU invariants of the three-qubit pure state.  

\section{Summary}
In this paper we have extended the mathematical framework of Ref.~\cite{supra2020} to explore SLOCC invariants of pure  three-qubit states. This method enables one to evaluate concurrences and tangle 
in terms of the {\em Lorentz invariant eigenvalues} of the $4\times 4$ real positive matrices $\Gamma_{ij}=G\,\Lambda_{ij}\, G\, \Lambda_{ij}^T,$ constructed from the real parametrizations $\Lambda_{ij}$ of the two-qubit subsystem density matrices $\rho_{ij},\ \ ij=AB, BC,CA$  of a pure state of three qubits. In particular, we have shown that (i) the matrices $\Gamma_{ij}$ evaluated in a pure three-qubit state have {\em at most} two distinct eigenvalues (i) the squared concurrence $C^2_{ij}$ is equal to the  least  eigenvalue of $\Lambda_{ij}$ and (ii) the three-tangle $\tau$ is equal to the difference between the highest and the smallest  eigenvalues of $\Gamma_{ij}$. This is illustrated in the example of permutation symmetric three-qubit pure states. Finally, we have given a set of {\em five} SLOCC invariants $\{ {\cal K}_1,\, {\cal K}_2,\, {\cal K}_3,\, {\cal K}_4,\,{\cal K}_5\}$, which are the natural algebraic generalizations of Ac{\'i}n's LU invariants $\{ I_1,\, I_2,\, 
I_3,\, I_4,\,I_5\}$.

\section*{Acknowledgements}
ARU, Sudha and BNK  are supported by the Department of Science and Technology (DST), India through Project No. DST/ICPS/QUST/2018/107. ASH is supported
by the Foundation for Polish Science (IRAP Project, ICTQT, contract no. MAB/2018/5, co-financed by EU within Smart Growth Operational Programme). HSK is supported by the Institute of Information \& Communications Technology Planning \& 14 Evaluation (IITP) Grant funded by the Korean government (MSIT) (No.2022-0-00463, Development of a quantum repeater in optical fiber networks for quantum internet).


\begin{thebibliography}{99}
\bibitem{fn1} It may be noted that $\sigma_2\, \sigma_\alpha\,\sigma_2= \displaystyle\sum_{\alpha'=0}^3\, g_{\alpha\alpha'}\, \sigma_\alpha'$, where $g_{\alpha\alpha'}$ are the elements of the  Minkowski metric $G={\rm diag}(1,-1,-1,-1)=G^T.$ Thus, $\displaystyle\sum_{\alpha,\beta=0}^3\, \Lambda_{\alpha\,\beta}\ (\sigma_2\, \sigma_\alpha\,\sigma_2)\otimes (\sigma_2\, \sigma_\beta\,\sigma_2) = \displaystyle\sum_{\alpha,\beta,\alpha',\beta'=0}^3 \,  \Lambda_{\alpha\,\beta}\,g_{\alpha\alpha'}\, g_{\beta\beta'}=\displaystyle\sum_{\alpha',\beta'=0}^3 \left(G\,\Lambda\,G\right)_{\alpha'\beta'}\, \sigma_\alpha'\otimes \sigma_\beta'$.
\end{thebibliography}
\end{document}